\documentclass[12pt]{article}
 \input epsf.sty

\begin{document}

\def\ket{\rangle}
\def\bra{\langle}
\def\CA{{\cal A}}
\def\CB{{\cal B}}
\def\CC{{\cal C}}
\def\CD{{\cal D}}
\def\CE{{\cal E}}
\def\CF{{\cal F}}
\def\CG{{\cal G}}
\def\CH{{\cal H}}
\def\CI{{\cal I}}
\def\CJ{{\cal J}}
\def\CK{{\cal K}}
\def\CL{{\cal L}}
\def\CM{{\cal M}}
\def\CN{{\cal N}}
\def\CO{{\cal O}}
\def\CP{{\cal P}}
\def\CQ{{\cal Q}}
\def\CR{{\cal R}}
\def\CS{{\cal S}}
\def\CT{{\cal T}}
\def\CU{{\cal U}}
\def\CV{{\cal V}}
\def\CW{{\cal W}}
\def\CX{{\cal X}}
\def\CY{{\cal Y}}
\def\CZ{{\cal Z}}

\newcommand{\todo}[1]{{\em \small {#1}}\marginpar{$\Longleftarrow$}}
\newcommand{\labell}[1]{\label{#1}}
\newcommand{\bbibitem}[1]{\bibitem{#1}\marginpar{#1}}
\newcommand{\llabel}[1]{\label{#1}\marginpar{#1}}
\newcommand{\dslash}[0]{\slash{\hspace{-0.23cm}}\partial}

\newcommand{\sphere}[0]{{\rm S}^3}
\newcommand{\su}[0]{{\rm SU(2)}}
\newcommand{\so}[0]{{\rm SO(4)}}
\newcommand{\bK}[0]{{\bf K}}
\newcommand{\bL}[0]{{\bf L}}
\newcommand{\bR}[0]{{\bf R}}
\newcommand{\tK}[0]{\tilde{K}}
\newcommand{\tL}[0]{\bar{L}}
\newcommand{\tR}[0]{\tilde{R}}

\newcommand{\btzm}[0]{BTZ$_{\rm M}$}
\newcommand{\ads}[1]{{\rm AdS}_{#1}}
\newcommand{\ds}[1]{{\rm dS}_{#1}}
\newcommand{\eds}[1]{{\rm EdS}_{#1}}
\newcommand{\sph}[1]{{\rm S}^{#1}}
\newcommand{\gn}[0]{G_N}
\newcommand{\SL}[0]{{\rm SL}(2,R)}
\newcommand{\cosm}[0]{R}
\newcommand{\hdim}[0]{\bar{h}}
\newcommand{\bw}[0]{\bar{w}}
\newcommand{\bz}[0]{\bar{z}}
\newcommand{\be}{\begin{equation}}
\newcommand{\ee}{\end{equation}}
\newcommand{\bea}{\begin{eqnarray}}
\newcommand{\eea}{\end{eqnarray}}
\newcommand{\pat}{\partial}
\newcommand{\lp}{\lambda_+}
\newcommand{\bx}{ {\bf x}}
\newcommand{\bk}{{\bf k}}
\newcommand{\bb}{{\bf b}}
\newcommand{\BB}{{\bf B}}
\newcommand{\tp}{\tilde{\phi}}
\hyphenation{Min-kow-ski}

\def\apr{\alpha'}
\def\str{{str}}
\def\lstr{\ell_\str}
\def\gstr{g_\str}
\def\Mstr{M_\str}
\def\lpl{\ell_{pl}}
\def\Mpl{M_{pl}}
\def\varep{\varepsilon}
\def\del{\nabla}
\def\grad{\nabla}
\def\tr{\hbox{tr}}
\def\perp{\bot}
\def\half{\frac{1}{2}}
\def\p{\partial}
\def\perp{\bot}
\def\eps{\epsilon}

\renewcommand{\thepage}{\arabic{page}}
\setcounter{page}{1}

\rightline{hep-th/0205311}
\rightline{UPR-997-T} \vskip 1cm

\centerline{\Large \bf Three point functions of  ${\cal N} = 4$
Super Yang Mills} \centerline{\Large \bf from Light Cone String
Field Theory in PP-wave} \vskip 0.5 cm
\renewcommand{\thefootnote}{\fnsymbol{footnote}}
\centerline{{ Min-xin Huang \footnote{minxin@sas.upenn.edu} }}
\vskip .5cm \centerline{\it David Rittenhouse Laboratories,
University of Pennsylvania} \centerline{\it Philadelphia, PA
19104, U.S.A.}

\setcounter{footnote}{0}
\renewcommand{\thefootnote}{\arabic{footnote}}

\begin{abstract}
In this paper we calculate three-string interaction from light
cone string field theory in pp-wave. We find exact agreements with
the free planar three point functions of non-chiral BMN operators
of ${\cal N} = 4$ super Yang Mills. The three string interaction
vertex involving the Neumann matrices was derived in a recent
paper hep-th/0204146. We explicitly calculate the bosonic Neumann
matrices in the limit of large $\mu p^{+} \alpha^{'}$ . Using the
Neumann matrices we are able to compute the cubic interactions of
three string modes in a pp-wave background.

\end{abstract}


\section{Introduction}
\label{intro} The AdS/CFT correspondence states that the ${\cal N}
= 4$ SU(N) Yang-Mills theory  is equivalent to IIB string theory
quantized on the $\ads{5} \times S^5$ background~\cite{jthroat}.
Recently Berenstein, Maldacena and Nastase \cite{BMN} have shown
that IIB superstring theory on a pp-wave background with
Ramond-Ramond flux is dual to a sector of ${\cal{N}}=4$ SU(N)
super Yang-Mills theory containing operators with large R-charge
$J$  . The pp-wave solution of type IIB supergravity has 32
supersymmetries and can be obtained as a Penrose limit of $AdS_5
\times S^5$ \cite{blau}. While the application of the AdS/CFT
correspondence in the usual $AdS \times S$ background is difficult
to go beyond the supergravity approximation on the string theory
side, the string worldsheet theory in the pp-wave background is
exactly solvable, as shown by \cite{metsaev}. More recently, there
have been some progress on the question of string interactions
\cite{seme, bn, gross, harvmit, spvo, gopakumar, kklp}.
\footnote{While we are preparing the manuscript the paper
\cite{kklp} appeared on the internet archive which partially
overlapped with results in section \ref {supergravity}. See Refs.
\cite{papa2} for related other recent developments.}

In \cite{harvmit} it is proposed that the matrix element for a
single string $|\Phi_3\rangle$ to split into a two-string state
$|\Phi_1\rangle|\Phi_2\rangle$ in the string field theory light
cone Hamiltonian is
\begin {equation}
\label{1}
(\Delta_3-\Delta_1-\Delta_2)\langle\bar{O}_3O_1O_2\rangle
\end {equation}
Where $O_i$'s are the properly normalized corresponding operators
in CFT and $\Delta_i$'s are their conformal dimensions. Here and
elsewhere in the paper we have omitted the factor of
$\frac{\Delta_{J_3,J_1+J_2}}{(4\pi
x_{12}^2)^{\frac{\Delta_1+\Delta_2-\Delta_3}{2}} (4\pi
x_{23}^2)^{\frac{\Delta_2+\Delta_3-\Delta_1}{2}}(4\pi
x_{13}^2)^{\frac{\Delta_1+\Delta_3-\Delta_2}{2}}}$ in the
correlator. In light cone string field theory the matrix element
is calculated by applying the three string states to a prefactor
$\hat{h}_3$ and the cubic interaction vertex state $|V\rangle$ in
the three-string Hilbert space ${\cal H}_3$. The $\hat{h}_3$ and
$|V\rangle$ are calculated in details in \cite{spvo}. It is
conjectured in \cite{harvmit} that the dressing factor
$(\Delta_3-\Delta_1-\Delta_2)$ in equation (\ref{1}) comes from
the prefactor $\hat{h}_3$ and assuming discretization of the
string world sheet at large $\mu$, a heuristic proof is given
there that the delta functional overlap agrees exactly with the
planar 3-point function in field theory, i.e.

\begin{equation}
\langle\Phi_1|\langle\Phi_2|\langle\Phi_3|V\rangle \sim
\langle\bar{O}_3O_1O_2\rangle
\end {equation}

In this paper we will explicitly check this proposal of
\cite{harvmit}. Here we will not calculate the overall
normalization of the matrix element. To make a explicit check of
the PP-wave/Yang-Mills duality, we will calculate the ratio with
vacuum amplitude on both sides. We should verify

\begin {equation}
\label{proposal}
\frac{\langle\Phi_1|\langle\Phi_2|\langle\Phi_3|V\rangle}{\langle
0_1|\langle 0_2|\langle 0_3|V\rangle}
=\frac{\langle\bar{O}_3O_1O_2\rangle}{\langle\bar{O}^J
O^{J_1}O^{J_2}\rangle}
\end {equation}
Here $J=J_1+J_2$, and $O^{J}=\frac{1}{\sqrt{N^JJ}}TrZ^J$ is the
properly normalized corresponding operator of the vacuum state.
The planar three point function of the vacuum states is

\begin{equation}\label{planarthh}
\langle \bar O^J(0)O^{J_1}(x_1)O^{J_2}(x_2) \rangle =
\frac{\sqrt{J_1J_2J}}{N}
\end{equation}

In \cite{harvmit} some free planar three point functions of BMN
operators are computed. Specifically, we have
\begin {equation}
\langle\bar{O}^J_{00}O^{J_1}_{0}O^{J_2}_{0}\rangle=\frac{J_1J_2}{N\sqrt{J}}
\end {equation}
\begin {equation}
\langle\bar{O}^J_{m}O^{J_1}_{0}O^{J_2}_{0}\rangle=-\frac{J^{3/2}}{N}\frac{\sin^2(\pi
mx)}{\pi^2m^2}
\end {equation}
\begin {equation}
\langle\bar{O}^J_{m}O^{J_1}_{n}O^{J_2}\rangle=\frac{J^{3/2}}{N}x^{\frac{3}{2}}\sqrt{1-x}\frac{\sin^2(\pi
mx)}{\pi^2 (mx-n)^2}
\end {equation}
Here $x=J_1/J$, and $O^{J_1}_{0}$, $O^{J_2}_{0}$ and $O^J_{00}$
are chiral operators

\begin {equation}
O^{J_1}_{0}=\frac{1}{\sqrt{N^{J_1+1}}} Tr(\phi^{I_1} Z^{J_1})
\end {equation}

\begin {equation}
O^{J_2}_{0}=\frac{1}{\sqrt{N^{J_2+1}}} Tr(\phi^{I_2} Z^{J_2})
\end {equation}

\begin {equation}
O^J_{00}=\frac{1}{\sqrt{N^{J+2}J}}\sum _{l=0}^{J}Tr(\phi^{I_1} Z^l
\phi^{I_2} Z^{J-l})
\end {equation}
$O^J_m$ is the BMN operator
\begin{equation}\label{bnnop}
O^J_{m} = \frac1{\sqrt{JN^{J+2}}} \sum_{l=0}^Je^{2\pi iml/J}
Tr(\phi^{I_1} Z^l\phi^{I_2} Z^{J-l}).
\end{equation}
So the ratios are
\begin {equation}\label{supergravityexample}
 \frac{\langle\bar{O}^J_{00}O^{J_1}_{0}O^{J_2}_{0}\rangle}{\langle
 \bar{O}^JO^{J_1}O^{J_2}\rangle}=\sqrt{x(1-x)}
 \end {equation}
\begin {equation}\label{stringexample1}
 \frac{\langle\bar{O}^J_{m}O^{J_1}_{0}O^{J_2}_{0}\rangle}{\langle
 \bar{O}^JO^{J_1}O^{J_2}\rangle}=-\frac{1}{\sqrt{x(1-x)}}\frac{\sin^2(\pi mx)}{\pi^2m^2}
 \end {equation}

 \begin {equation} \label{stringexample2}
 \frac{\langle\bar{O}^J_{m}O^{J_1}_{n}O^{J_2}\rangle}{\langle
 \bar{O}^JO^{J_1}O^{J_2}\rangle}=x\frac{\sin^2(\pi
mx)}{\pi^2 (mx-n)^2}
 \end {equation}

 In this paper we will calculate the ratios of the three point correlators (\ref{supergravityexample})
  (\ref{stringexample1}) (\ref{stringexample2}) from light cone string field theory in
  pp-wave. We will find exact agreements with equation (\ref{proposal}).
   The paper is organized as follows. In section \ref{supergravity}
we calculate the easier example of supergravity modes and verify
equation (\ref{supergravityexample}). In Section \ref{string} we
calculate the other two examples for the non-chiral BMN operators
and verify equation (\ref{stringexample1}) (\ref{stringexample2})
. The techniques in section \ref{string} are basically the same as
section \ref{supergravity}.

Some notation is the following. Following the notation of
\cite{spvo} we denote $\alpha=\alpha^{'}p^{+}$. The strings are
labeled by $r=1,2,3$ and in light-cone gauge their widths are $2
\pi|\alpha_{(r)}|$, with $\alpha_{(1)} + \alpha_{(2)} +
\alpha_{(3)} = 0$. We will take $\alpha_{(1)}$ and $\alpha_{(2)}$
positive for purposes of calculation. Also note that
\begin{equation}
x=\frac{J_1}{J}=\frac{|\alpha_{(1)}|}{|\alpha_{(3)}|}
\end{equation}

\section{ Interaction of supergravity modes}
\label{supergravity}

Light cone string field theory is an old subject dating back to
the 80's (see \cite{GreenXX}-\cite{GSB}). For the purpose of this
paper we will need to use the cubic interaction vertex
$|V\rangle$, which can be written as an element in the 3-string
Hilbert space. Roughly speaking, the interaction amplitude of
three strings is the inner product of the three string state with
the cubic interaction vertex.

The string modes interaction vertex is $|V\rangle=E_aE_b|0\rangle$
where $E_a$ and $E_b$ are bosonic and fermionic operators that are
calculated in details in \cite{spvo} . Here will not consider the
fermionic sector. Up to a overall factor, the bosonic operator
$E_a$ is

\begin{equation}
 E_a \sim \exp \left[  \frac{1}{2} \sum_{r,s = 1}^3\sum_{I =
 1}^8 \sum_{m,n=-\infty}^{\infty}
a_{m(r)}^{\dagger I} \bar{N}^{(rs)}_{(mn)} a_{n(s)}^{\dagger I}
\right]
\end{equation}
where $I$ denote the eight transverse directions, and the Neumann
matrices $\bar{N}^{(rs)}_{(mn)}$ are computed in the Appendix in
the limit of large $\mu p^{+} \alpha {'}$.

 We consider the interaction of three supergravity modes
 $a^{\dagger I_1}_{0(1)}|0\rangle$,$a^{\dagger I_2}_{0(2)}|0\rangle$,$a^{\dagger I_1}_{0(3)}a^{\dagger
 I_2}_{0(3)}|0\rangle$. Here $I_1$ and $I_2$ are two different transverse directions. We want to compute
 the object
\begin {equation}
 \frac{\langle0|a^{I_1}_{0(1)}a^{I_2}_{0(2)}a^{I_1}_{0(3)}a^{I_2}_{0(3)}|V\rangle}{\langle0|V\rangle}
 \end {equation}
We will need the zero components of the Neumann matrices. From
equation (\ref{Neumann}) we find
$\bar{N}^{(13)}_{(00)}=\bar{N}^{(31)}_{(00)}=-\sqrt{x}$,
$\bar{N}^{(23)}_{(00)}=\bar{N}^{(32)}_{(00)}=-\sqrt{1-x}$.
(Actually this is true without taking the large $\mu p^{+}
\alpha^{'}$ limit.)

 From the Baker-Hausdorff formula
\footnote{The Baker-Hausdorff formula is
$e^{A}Be^{-A}=B+[A,B]+\frac{1}{2!}[A,[A,B]]+\cdots$.} we know
\begin{equation}
 (E_a)^{-1}a^{I_1}_{0(1)}a^{I_1}_{0(3)}E_a=
 a^{I_1}_{0(1)}a^{I_1}_{0(3)}-\frac{1}{2}(\bar{N}^{(13)}_{(00)}+\bar{N}^{(31)}_{(00)})
 \end {equation}
 \begin{equation}
 (E_a)^{-1}a^{I_2}_{0(2)}a^{I_2}_{0(3)}E_a=
 a^{I_2}_{0(2)}a^{I_2}_{0(3)}-\frac{1}{2}(\bar{N}^{(23)}_{(00)}+\bar{N}^{(32)}_{(00)})
 \end {equation}
So
\begin {equation} \label{ppsupergravityexample}
 \frac{\langle0|a^{I_1}_{0(1)}a^{I_2}_{0(2)}a^{I_1}_{0(3)}a^{I_2}_{0(3)}|V\rangle}{\langle0|V\rangle}
 =\frac{1}{2}(\bar{N}^{(13)}_{(00)}+\bar{N}^{(31)}_{(00)})\frac{1}{2}(\bar{N}^{(23)}_{(00)}+\bar{N}^{(32)}_{(00)})=\sqrt{x(1-x)}
 \end {equation}
On the field theory side, the three modes $a^{\dagger
I_1}_{0(1)}|0\rangle$,$a^{\dagger
I_2}_{0(2)}|0\rangle$,$a^{\dagger I_1}_{0(3)}a^{\dagger
I_2}_{0(3)}|0\rangle$ correspond to chiral operators $O^{J_1}_0$,
$O^{J_2}_0$ and $O^{J}_{00}$ (Suppose $I_1$ and $I_2$ correspond
to scalar instead of the $D_\mu$ insertions in the string of
$Z$'s) . Thus we have found equation (\ref{ppsupergravityexample})
is in agreement with equation (\ref{supergravityexample}).

\section{Interaction of string theory modes}
\label{string}

\subsection{Example 1}
We consider the interaction of three states
$a^{I_1(BMN)\dagger}_{0(1)}|0\rangle$,$a^{I_2(BMN)\dagger}_{0(2)}|0\rangle$,
$a^{I_1(BMN)\dagger}_{m(3)}a^{I_2(BMN)\dagger}_{-m(3)}|0\rangle$,
which correspond to operators $O^{J_1}_{0}$, $ O^{J_2}_{0}$ and
$O^J_{m}$. We caution the reader here the $a^{+}$ notation we use
 is not the familiar string theory basis of BMN \cite {BMN}, but is the
same as in \cite{spvo}. These two basis are related by
\begin {equation}
a^{BMN}_n=\frac{1}{\sqrt{2}}(a_{|n|}-ie(n)a_{-|n|})
\end {equation}
where $e(n)$ is the sign of $n$ (For $n=0$, $a^{BMN}_0=a_0$).
Notice that the $a_{-n}$ mode contribution vanish since the
corresponding Neumann matrices elements are zero. The calculation
here follows similarly as in section \ref{supergravity}, we find
\begin{eqnarray}
&&\frac{\langle 0|a^{I_1(BMN)}_{0(1)} a^{I_2(BMN)}_{0(2)}
a^{I_1(BMN)}_{m(3)}a^{I_2(BMN)}_{-m(3)}|V\rangle}{\langle0|V\rangle}
\nonumber
\\&& =\frac{\langle 0|a^{I_1}_{0(1)} a^{I_2}_{0(2)}
a^{I_1}_{m(3)}a^{I_2}_{m(3)}|V\rangle}{2\langle0|V\rangle}\\&&
=\frac{1}{2}\bar{N}^{(31)}_{(m0)}\bar{N}^{(32)}_{(m0)} \nonumber
\end{eqnarray}
Using equation (\ref{Neumann}) we find
$\frac{1}{2}\bar{N}^{(31)}_{(m0)}\bar{N}^{(32)}_{(m0)}=-\frac{1}{\sqrt{x(1-x)}}\frac{\sin^2(\pi
mx)}{\pi^2m^2}$, in agreement with equation
(\ref{stringexample1}).

\subsection{Example 2}

In this example we consider the interaction of three states
$a^{I_1(BMN)\dagger}_{n(1)}a^{I_2(BMN)\dagger}_{-n(1)}|0\rangle$,$|0\rangle$,
$a^{I_1(BMN)\dagger}_{m(3)}a^{I_2(BMN)\dagger}_{-m(3)}|0\rangle$,
which correspond to operators $O^{J_1}_{n}$, $ O^{J_2}$ and
$O^J_{m}$. Notice the Neumann matrix elements $\bar{N}^{(rs)}_{(m
,-n)}$ and $\bar{N}^{(rs)}_{(-m,n)}$ vanish,
 so
\begin{eqnarray}
&&\frac{\langle 0|a^{I_1(BMN)}_{n(1)} a^{I_2(BMN)}_{-n(1)}
a^{I_1(BMN)}_{m(3)}a^{I_2(BMN)}_{-m(3)}|V\rangle}{\langle0|V\rangle}
\nonumber
\\&&
=\frac{1}{4}(\bar{N}^{(31)}_{(m,n)}-\bar{N}^{(31)}_{(-m,-n)})^2
\\&& =x\frac{\sin^2(\pi mx)}{\pi^2 (mx-n)^2}\nonumber
\end{eqnarray}
Again it agrees with equation (\ref{stringexample2}).

\section{Conclusion}
In this paper we explicitly check the pp-wave/Yang-Mills duality
for  BMN operators in free planar limit. By retaining sub-leading
order in the Neumann matrices, we can develop a systematic
expansion of light cone string field theory in $1/(\mu
p^{+}\alpha{'})$. It would be interesting to check equation
(\ref{proposal}) in sub-leading order of $1/(\mu p^{+}\alpha{'})$
expansion by turning on interaction in the three point functions.

\vspace{0.2in} {\leftline {\bf Acknowledgments}}

We are grateful to Vijay Balasubramanian, Thomas S. Levi and Asad
Naqvi for illuminating discussions and critical readings of the
manuscript.

\appendix

\section{Computation of the bosonic Neumann matrices in large $\mu p^{+} \alpha^{'}$ limit}
In this appendix we will show that the infinite dimensional
Neumann matrices turn out to simplify in large $\mu p^{+}
\alpha^{'}$ limit. The Neumann matrices $\bar{N}^{(rs)}_{mn}$
($r,s=1\cdots 3, m,n=-\infty \cdots +\infty$ ) is calculated in
\cite{spvo}

\begin{equation}
 \overline{N}_{mn}^{(rs)} = \delta^{rs} \delta_{mn} - 2 \sqrt{
\omega_{m(r)} \omega_{n(s)} } (X^{(r) {\rm T}} \Gamma_a^{-1}
X^{(s)})_{mn}
\end{equation}

where $\omega_{m(r)}=\sqrt{m^2+(\mu \alpha_{(r)})^2}$, and

\begin{equation} \label{gamma}
(\Gamma_a)_{mn} = \sum_{r=1}^3 \sum_{p=-\infty}^\infty
\omega_{p(r)} X^{(r)}_{mp} X^{(r)}_{np}
\end{equation}

The definition of $X^{(r)}$ is the following. Consider for $m,n>0$
the matrices of \cite{GreenTC, spvo},
\begin{equation}
 A^{(1)}_{mn}  =(-1)^{n} {2 \sqrt{m n} \over \pi} {x \sin{m
\pi x}\over n^2-m^2 x^2},
\end {equation}
\begin{equation}
A^{(2)}_{mn} =- {2 \sqrt{m n} \over \pi} {(1-x) \sin{m \pi x}
\over n^2-m^2 (1-x)^2}
\end{equation}

\begin{equation}
C_{mn} = m \delta_{mn}
\end{equation}
 and the vector
\begin{equation}
 B_m = - {2 \over \pi} {\alpha_{(3)} \over
\alpha_{(1)} \alpha_{(2)}} m^{-3/2} \sin m \pi x
\end{equation}
 We define
$X^{(3)}_{mn} = \delta_{mn}$, while for $r=1,2$ we can express the
matrices $X^{(r)}$ as

\begin{eqnarray}
 X^{(r)}_{mn} &=& (C^{1/2} A^{(r)}
C^{-1/2})_{mn} \qquad\qquad\qquad{\rm if}~m,n>0,\nonumber \\& =&
{\alpha_{(3)} \over \alpha_{(r)}} (C^{-1/2} A^{(r)} C^{1/2})_{
-m,-n}, \qquad{\rm if}~m,n<0,\nonumber \\&= & - {1 \over \sqrt{2}}
\epsilon^{rs} \alpha_{(s)} (C^{1/2} B)_m\qquad \qquad~~~{\rm
if}~n=0~{\rm and}~m>0,\\
&=&1\qquad\qquad\qquad\qquad\qquad\qquad\qquad{\rm if}~m=n=0,
\nonumber \\&=&0\qquad\qquad\qquad\qquad\qquad\qquad\qquad{\rm
otherwise}. \nonumber
\end{eqnarray}

In the limit of large $\mu\alpha$, $\omega_{m(r)}=\sqrt{m^2+(\mu
\alpha_{(r)})^2}\approx \mu|\alpha_{(r)}|$. Using equation
(\ref{gamma}), we find that for $m,n> 0$,
\begin {eqnarray}
(\Gamma_a)_{mn}&=& |\alpha_{(3)}|\mu\frac{4mn}{\pi^2}\sin(m\pi
x)\sin(n \pi x)
[\sum_{l=1}^{+\infty}\frac{x^3}{(l^2-m^2x^2)(l^2-n^2x^2)}\\&&
+\sum_{l=1}^{+\infty}\frac{(1-x)^3}{(l^2-m^2(1-x)^2)(l^2-n^2(1-x)^2)}
 +\frac{1}{2m^2n^2x(1-x)}]+|\alpha_{(3)}|\mu\delta_{mn} \nonumber
\end {eqnarray}

\begin {eqnarray} (\Gamma_a)_{-m,-n}&=&
|\alpha_{(3)}|\mu\frac{4}{\pi^2}\sin(m\pi x)\sin(n \pi x)
[\sum_{l=1}^{+\infty}\frac{xl^2}{(l^2-m^2x^2)(l^2-n^2x^2)} \\&&+
\sum_{l=1}^{+\infty}\frac{(1-x)l^2}{(l^2-m^2(1-x)^2)(l^2-n^2(1-x)^2)}
 ]+|\alpha_{(3)}|\mu\delta_{mn} \nonumber
\end {eqnarray}
 and $(\Gamma_a)_{00}=2|\alpha_{(3)}|\mu$. All other components
 such as $(\Gamma_a)_{m0}$ are zero.

Using the summation formulae in the appendix D of \cite{GreenTC}
we find
\begin{equation}
(\Gamma_a)_{mn}=2|\alpha_{(3)}|\mu\delta_{mn}\qquad\qquad\qquad\qquad
for \qquad m,n=-\infty\cdots +\infty
\end {equation}
So the Neumann matrices in large $\mu|\alpha|$ limit is
\begin{equation} \label {Neumann}
\bar{N}^{(rs)}_{(mn)}=\delta^{rs}\delta_{mn}-\frac{\sqrt{|\alpha_{(r)}||\alpha_{(s)}|}}{|\alpha_{(3)}|}(X^{(r)T}X^{(s)})_{mn}
\end{equation}
We note $\bar{N}^{(rs)}_{(mn)}=\bar{N}^{(sr)}_{(nm)}$.

In \cite{spvo} the cubic coupling matrix of supergravity modes is
derived. It is

\begin{equation} \label{supergravityNeumann}
M^{rs} = \left(\matrix{ 1-x &-\sqrt{x(1-x)} &-\sqrt{x} \cr
-\sqrt{x(1-x)} &x &-\sqrt{1-x} \cr -\sqrt{x} &-\sqrt{1-x} & 0 }
\right)
\end{equation}
One would be tempted to identify $M^{rs}$ as the zero-zero
component of the Neumann matrices (\ref{Neumann}). But this is
incorrect.\footnote{We thank M. Spradlin  and A. Volovich for
pointing out this to us.} Actually one can check
$M^{rs}=\bar{N}^{(rs)}_{(00)}$ is true when $\mu p^{+}
\alpha^{'}=0$, but at large $\mu p^{+} \alpha^{'}$ limit
$M^{rs}=\bar{N}^{(rs)}_{(00)}$ is true only when $r=3$ or $s=3$.

\end{document}